\documentclass[12pt,a4paper]{article}          
\usepackage{graphicx}
\advance\textheight1cm
\advance\topmargin-1cm
\advance\oddsidemargin-0.5cm
\advance\evensidemargin-0.5cm
\advance\textwidth1cm
\begin{document}
\title{Primacy analysis of  the system of Bulgarian cities}
\author{Zlatinka I. Dimitrova$^1$, Marcel Ausloos$^2$}
\date{}
\maketitle
\begin{abstract}
We study the primacy in the Bulgarian urban system. Two groups of cities are 
studied: (i) the  whole  Bulgaria city system
that contains about 250 cities and is studied in the time interval between 2004 and 2011; and (ii)    
A system of 33 cities,  studied over the time interval 1887 
till 2010. For these cities the 1946 population was over $10\ 000$ 
inhabitants.  The notion of  primacy in the two systems of cities is
studied  first  from the global primacy index of Sheppard [$^1$]. Several  (new) additional indices 
are introduced in order   to compensate defects in the Sheppard index.  Numerical 
illustrations are illuminating through the so called "length ratio".\\
{\bf Key words:} city sizes, Zipf's law,  primacy, primacy indices\\
{\bf PACS:} 89.75.Da. Systems obeying scaling laws;  89.75.Fb. Structures and organization in 
complex systems
\end{abstract}
\section{Introduction}\label{intro}
Nonlinearity [$^{2-5}$] and complexity [$^{6-9}$] are common features of a  
large number of systems studied in  modern science [$^{10-12}$]. Such systems
are much investigated by  nonlinear dynamics  methods,
and time series analysis [$^{13-17}$]. In the last decade or so, these methods
have been applied also to many social, economic, and financial systems 
[$^{18-20}$].
In many cases,  researchers have detected the existence of power laws, 
for different characteristic quantities of these complex systems. 
Power laws are  useful tools in studying complex systems because scaling relations 
may indicate that the system is controlled by a few rules that propagate across a 
wide range of scales [$^{21,22}$]. 
\par 
Below we analyze data from a specific
nonlinear complex system where power laws can be observed: the city population system of a  
specific country.  In the course of  time, the cities in a country  develop a  
hierarchy. An expression of this hierarchy is the city population size distribution that can 
be easily   constructed for any urban system. Zipf [$^{23}$]  suggested that a large number 
of observed city population  size distributions could be approximated by a simple  {\it 
scaling (power) law} $N_r = N_1/r $,   where $N_r$ is the population of the $r$-th largest 
city. A more flexible equation, with two parameters,  reads
$N_r = {N_1}/{r^\beta}$,
is called the rank-size scaling law. Zipf suggested that the particular case  $\beta=1$ 
represents a desirable situation, in which   forces of concentration
balance those of decentralization. Such a case is called  the rank-size rule. 
The  urban population size distribution  of  developed countries, like the USA,  fits very 
well the rank-size rule over several decades [$^{24,25}$]. 
\par
In this paper we discuss the human population of Bulgaria.
In Bulgaria exist about  250 cities and about 4000 villages. 
The human population of the country   reached almost 9 million in 1985 but after this it has decreased 
steadily  in the last 25 years reaching 7.3 million in 2011. Below, we examine  two sets of  urban  
population data. The first set  is the yearly count of the population of  whole Bulgarian
cities  from 2004 till 2011,  as  recorded by the National Statistical Institute
of the Republic of Bulgaria  ($http: www.nsi.bg$). The second data set is the yearly population count  in 
1887, 1910, 1934, 1946, 2000 and 2011  for  the 33 Bulgarian cities  which had a population
 over $10 \ 000$ citizens in 1946. The data for   1887,
1910, 1934, 1946 is taken from  from [$^{26}$]  while  the data from 2000 and 2010 are from
the National Statistical Institute of Republic of Bulgaria.
\section{Analysis of primacy}\label{sect2}
An important problem connected to the real city size distributions is the 
problem of primacy. It has been seen that, in a few cases,  the city size distributions 
can be close to the rank-size relationship of Zipf. In  most cases 
these distributions are primate  distributions $[^1]$, i.e. 
when one or  very few but very large cities (the capital and several other 
cities) predominate  the distribution; convex distributions that 
correspond to presence of number of large cities ; or distributions
with some mix of primacy and convexity, leading to S-shape like or more 
complicated structure.
\par 
Measures of primacy can be of the kind
\begin{equation} \label{a4}
Pr^{(k)} = \frac{N_1}{\sum_{r=2}^k N_r}, k=2,3,\dots .
\end{equation}
Eq.(\ref{a4}) gives a numerical value for the primacy of the largest city
with respect to the next $k-1$ cities if the cities are ordered by decreasing
number of inhabitants. If a power law of the kind $N_r =  N_1/r^\beta$ is substituted into
each of these measures, it is obvious  that the corresponding 
index of primacy depends on $\beta$,  whence 
the rank-size relationships with different slopes will have 
different levels of primacy.  Then, it will be not possible to discriminate
between a country where a primate city dominates a city size
distribution, which otherwise may have a low and fairly consistent 
negative slope, from a country exhibiting a rank-size
relationship with  steep slope $\beta$.  Sheppard $[^1]$ tried to avoid this puzzle
by formulating  a {\it primacy index} that is independent of $\beta$, i.e.,  he defined
\begin{equation} \label{a7}
Pr_N =\frac{1}{N-2} \sum_{r=1}^{N-2} \bigg[ \frac{\ln(N_r+1)-\ln(N_{r})}{\ln(N_{r+2}) - \ln(N_{r+1}))} \bigg] 
\bigg[ \frac{\ln(r+2)-\ln(r+1)}{\ln(r+1)-\ln(r)} \bigg]
\end{equation}
The logics behind  this index is as follows. Let us substitute here the power law
rank-size relationship $N_r = N_1 r^{-\beta}$. The result is $
Pr_N=(1/(N-2)) \sum_{r=1}^{N-2} 1 = 1$.
Thus, for a perfect power law rank-size relationship, the index $Pr_N$ has a value of
1, irrespective of the slope of the relationship.
\par
We have applied  the Sheppard index, Eq.(\ref{a7}), to study the primacy (or 
"hierarchy")  of Bulgarian cities in the years between
2004 and 2011. 
\begin{figure}[t]
\begin{center}
\includegraphics[scale=0.9]{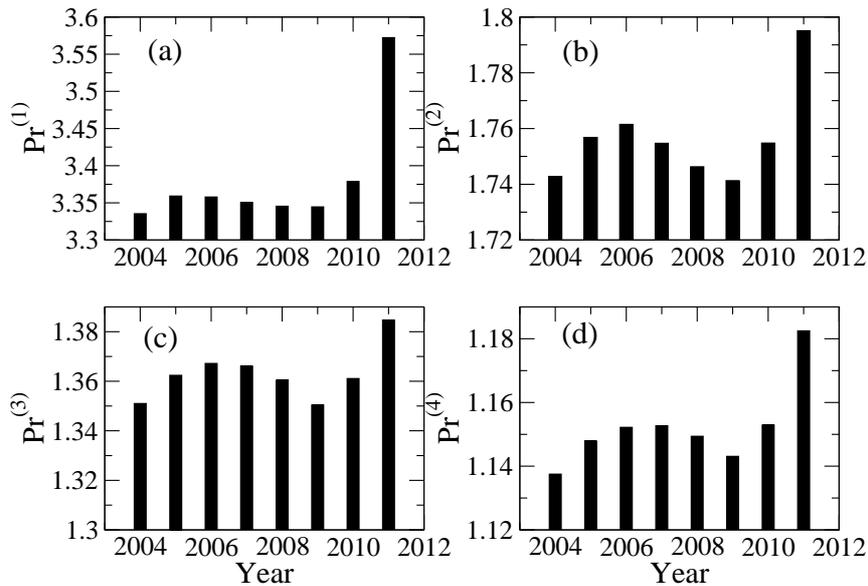}
\end{center}
\caption{Evolution of the first 4 primacy indices for Sofia,
 the capital of Bulgaria,  from 2004 till 2011. Remember 
that $Pr^{(k)} = N_1/(\sum_{i=2}^k N_k)$, for $k=2,3,\dots$, Eq.(\ref{a4}).
Till 2006, the values of the primacy indices increase; later, 
  till 2009, the values of the indices decrease. Next,  a
relatively sharp increase is observed in 2010 and 2011.}
\end{figure}
Figure 1 shows the changes in the first 4 primacy indices $Pr^{(1)}$,  ...,  $Pr^{(4)}$ for the largest
city (and capital) of Bulgaria: Sofia. A decreasing of primacy is observed between
2006 and 2009. One reason for this is economic: the good economic development
before the crisis (that appeared in Bulgaria in 2009). Because of 
favorable economic conditions, there was enough inflow of people to the second,
third, and the fourth largest city, thereby  decreasing  the primacy of the capital,
Sofia. However the subsequent economic crisis worsened the job perspectives in the above mentioned large
cities which led to an increased  inflow of people  back to Sofia. This led to increasing the 
primacy of the capital in the last few  years.
\par
Figure 2 (a)-(c) shows the evolution of the population of the capital Sofia within the
class of the 33 cities  (with population exceeding  $10 \ 000$ in 1946). It is  
seen that
the primacy in 1887 was below 1. Note that Sofia was the capital but not the largest
city in Bulgaria up to 1890. Figure 6 (d) shows how the advantage to be a  capital
was favorable for the population increase.  In several more words, in 1879,  
almost an year after creation of the Third Bulgarian state a new capital had
to be selected. There were two canditate cities: Sofia and Veliko Tarnovo (the capital
of the Second Bulgarian state). Sofia was selected  to be the capital of Bulgaria (Sofia
won by 1 vote over  the old capital Veliko Tarnovo).
At this time,  the population of Sofia was about  twice  larger than the population
of Veliko Tarnovo. The concentration process  led to a situation in which the
population of Sofia became 25 times larger than the population of Veliko Tarnovo. 
In the last 25 years, the total country population as well as the urban population have 
decreased but the population of these two cities has further  increased: and the
rate of increase of the  Veliko Tarnovo  population is larger that the rate of
increasing  rate of Sofia population. Thus in 2010 the population of Sofia is about
15 times larger than the population of Veliko Tarnovo. 
This is a strong  evidence for the
fact that the population growth of the Bulgarian cities is size dependent. 
 \begin{figure}[t]
\begin{center}
\includegraphics[scale=0.9]{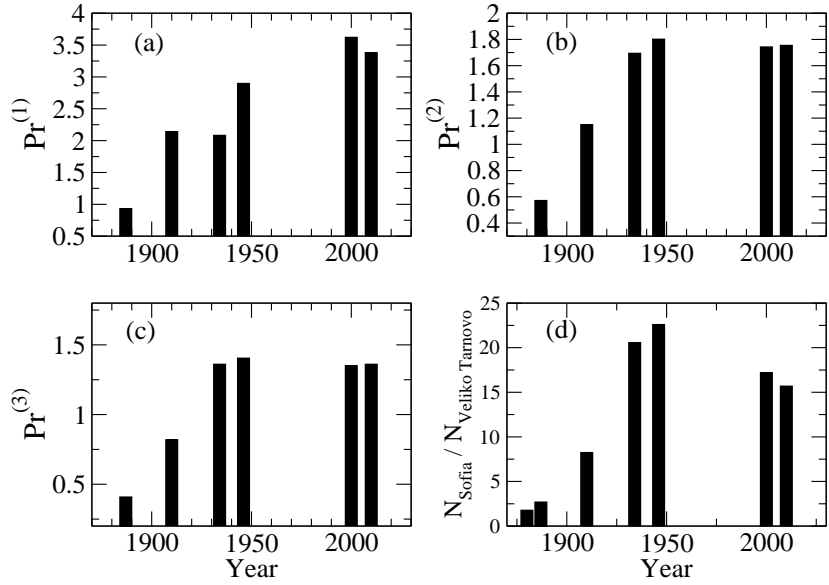}
\end{center}
\caption{ (a), (b), (c): primacy of Sofia in the system of 33 cities for 1887,
1910, 1934, 1946, 2000, and 2010; (d): on the advantage to be a capital.
Ratio of the population of the two cities that were candidates for capital
of Bulgaria in 1879. The ratios are for 1880, 1887, 1910, 1934, 1946, 2000,
and 2010.}
\end{figure}
\par 
The primacy index of Sheppard contains a variance of two logarithms in the
denominator. When two cities have almost the same number
of citizens this variance can be very small thus leading to large value of
the Sheppard index. Actually, this happened:  when we analyzed primacy in the (large)
system of about 250 Bulgarian cities;  there were two cities for which number of
citizens  differs from each other by 1 only. In order to avoid 
such a kind of problems we propose  to consider two  other local primacy measures where the variance 
of logarithms is present only in the numerator, as follows.
Let  the cities be ranked, in the  each
studied city system,  according to the   population, i.e. $N_{r} \ge N_{r+1}$. The measures are:
\begin{equation} \label{v1}
V_{r}=\frac{\ln(N_{r-1})-\ln(N_r)}{\ln(r) -  \ln (r-1)}
\end{equation}
and 
\begin{equation}\label{v2}
W_{r} =   \frac{\ln(N_r) - \ln(N_{r+1})}{
\ln(r+1) - \ln(r)} -  \frac{\ln(N_{r+1}) - \ln(N_{r+2})}{
\ln(r+2) - \ln(r+1)} \equiv V_{r+1} - V_{r+2}
\end{equation}
For the case of power law relationship $N_r = N_1 r^{-\beta}$ ($r=1,\dots,N$) for
each $\beta \ge 0$ the values of the measures $V_{i}$ and $W_{i}$ are as follows: 
$V_{r} = \beta$ ($r=2,\dots,N-1$) and $W_{r} = 0$ ($r=2,\dots,N-2$).
\par 
Let us now discuss cases when deviations from the power law occur. Let us consider the $(\ln (r), 
\ln(N_r))$-plane.  Suppose first that $\ln (N_{r-1})$ is fund over the straight line formed 
between the points $\ln (N_r)$ and 
$\ln (N_{r+1})$, a case of {\it local primacy}. In such a  case, $V_{r} > V_{r+1}$
and $W_{r-1}>0$. However, if $\ln(N_{r-1})$ is below the straight line formed by the points $\ln (N_r)$ 
and $\ln (N_{r+1})$, a case of  {\it local convexity}, then  $V_{r} < V_{r+1}$ and $W_{r-1}<0$. 
Of course, if 
$\ln (N_{r-1})$ lies on the straight line formed by the points $\ln (N_r)$ and 
$\ln (N_{r+1})$, i.e. the power law case, then $V_{r} = V_{r+1}$ and $W_{r-1}=0$. 
Thus for the power law case $V$ will form a straight line as a function of $r$ and the
deviation of $V$ from a straight line will be a signal for
deviation of city size distribution from a power law function. 
\begin{figure}[t]
\begin{center}
\includegraphics[scale=0.8]{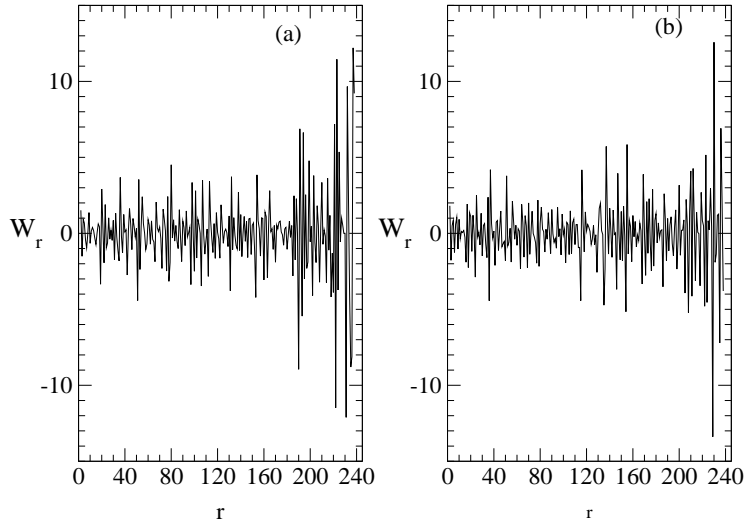}
\end{center}
\caption{W-measures for the system of Bulgarian cities: (a) 2004;  (b) 
2011 }
\end{figure}
\par
Fig.3 shows the $W$-measures for the system of all Bulgarian cities  in  2004 and 2011.
If a single power law was present in the rank-size relationship then $W_{r} = 0$ and $W_{r}$ would  be a straight line as a function of $r$. As easily observed, this is not the case for the system of
Bulgarian cities, since what is observed in the Fig. 3  is a mix of regions of local primacy and regions of local convexity. 
      \begin{table}\label{RN}
      \begin{tabular}{|c|c|c|c|c|c|c|c|c|c| }

\hline     &year:& 2004 & 2005 & 2006 & 2007 & 2008 & 2009 & 2010 & 2011  \\ 
\hline \hline
N:& &&&&&&&&  \\
\hline
50 &&  2.1171 & 2.2677 & 2.1652 & 2.2441 & 2.3363 & 2.3823 & 2.4169 & 2.4442  \\
\hline
100&& 2.3458 & 2.3564 & 2.2130 & 2.1390 & 2.2488 & 2.3856 & 2.2360 & 2.2755  \\
\hline
150&& 2.3987& 2.5607& 2.3736& 2.3122& 2.4431& 2.7221 & 2.4188 & 2.3798  \\
\hline
240&& 3.2635& 3.5600& 2.9865& 2.7529& 3.2125& 3.4156& 3.2255& 3.0221  \\
\hline
\end{tabular} 
\caption{Length ratio $R_N$ of the $V$-measure line for different number $N$  of  BG cities, ranking from $r= 1$ till $N$, 
from 2004 till 2011.}%
\end{table}
In order to characterize the deviation of the system of cities from a
system with the same number of cities, but obeying a power law, one  can
e.g. measure the length of the curves corresponding to the $V$-measures. 
Indeed, let us  consider  $N$ cities. If
the rank-size distribution of these cities is  a single power law, then
the $W$-measure of each 3  neighboring cities is equal to  $0$. For
a system of $N$ cities, there will be $N-2$ points in the ($r, W_r$) plane 
with coordinates $(j,0)$ where $j=1,\dots, N-2$. These $N-2$ points connect
$N-3$ segments of the $W$-curve and each segment has the same length $1$. (Remember
that we discuss the case when a power law holds for the distribution of cities populations).
Then, the  total length of the $W_r$ line in the ($r, W_r$)-plane  is  
$L_{\beta}=N-3$.
\par
Let now consider the case when the distribution of the cities population does $not$ behave  according to a power law.
Then, the $W_r$ curve is not a straight line (see Fig. 3 for an example); the
length of such a curve is  bigger than $L_{\beta}$. Next, let us define the {\it length ratio}
\begin{equation}\label{v4}
R_N = \frac{L_N}{L_{\beta}}
\end{equation} 
where $L_N$ is the length of the line associated with the corresponding $W_{r}$-index:
\begin{equation}\label{v5}
L_N = \sum_{r=1}^{N-3}\sqrt{1+(W_{r+1}-W_{r})^2}
\end{equation}  
The results for the length ratio $R_N$ for several classes of Bulgarian
cities are shown in Table 1.  For a given  number of cities,   
 the evolution of the deviation of the city size distribution
from a   power law\footnote{ for a single power law $R_N=1$} can be calculated. 
For   example, for the 50 largest cities,
$R_{50}$ increases steadily since 2006. This means that the populations of cities change
  in such a manner that the corresponding rank-size
distribution deviates more and more from a single power law. 
The
evolution  with respect to $R_N$  of the 100 largest cities is even more interesting,   since between
2006 and 2008  the distribution  appears to be  more like a single 
power law than for  2005. 
\par
Finally, note that the length ratio $R_N$ can be generalized in order to investigate the distribution 
deviation from a power law for any sub-class of cities, e.g. ranking between $N_1$
and $N_2$. One can define, e.g., 
\begin{equation}\label{v6}
R_{N_1,N_2} = \frac{1}{L_{\beta}} \sum_{r=N_1}^{N_2-3}\sqrt{1+(W_{r+1}-W_{r})^2}
\end{equation}  
\par
As concluding remark we note the following.
In this paper,   the city primacy  has been investigated for the two so defined 
groups of Bulgarian cities on the basis of the conventional  index of
Sheppard. However, we have indicated that other measures should be useful and have given 
definitions and subsequent numerical results. In particular we have defined and 
discussed results obtained by the measures $V_r$, $W_r$, and $R_N$. 
These measures
can be used to quantify the deviation of the rank-size distribution of a system
of cities from a power-law rank-size relationship. These measures can be applied not only for
a group of cities but also for any any group of objects that can be ranked on the
basis of some quantitative characteristics. 
\begin{flushleft}{\bf Acknowledgement}
\end{flushleft}
{\small This work has been performed in the framework of COST Action IS1104 
"The EU in the new economic complex geography: models, tools and policy evaluation".
We acknowledge some support
through the project 'Evolution spatiale et temporelle d'infrastructures r\'egionales 
et \' economiques en Bulgarie et en F{\'e}d{\'e}ration Wallonie-Bruxelles' within the  
intergovernemental agreement for cooperation between the Republic of  Bulgaria and la Communaut\' e 
Fran\c{c}aise de Belgique.}
\begin{flushleft}
{\bf REFRENCES}
\end{flushleft}         
$[^1]$ Sheppard E., City size distributions and spatial economic change.
WP-82-31, Working papers of the International Institute for
Applied System Analysis, Laxenburg, Austria, 1982.\\
$[^2]$ Kantz H., T. Schreiber, Nonlinear time series analysis.
Cambridge University Press, Cambridge, 1997.\\
$[^3]$ Vitanov N. K.,  Phys. Rev. E {\bf 62}, 2000, No.3, 3581 - 3591.\\
$[^4]$ Vitanov N. K., F. H. Busse, ZAMP {\bf 48}, 1997, No.2, 310-324. \\
$[^5]$ Martinov N., N. Vitanov, J. Phys. A: Math. Gen {\bf 25}, 1992,
L419-L425.\\
$[^6]$ Axelrod R., M. D. Cohen, Harnessing complexity: Organizational
implications on a scientific frontier. Free Press, New York, 1999.\\
$[^7]$ Boeck T.,  N. K. Vitanov,  (2002). Phys. Rev. E {\bf 65}, 2002, 
No. 3, Article number: 037203.\\ 
$[^8]$ Ashenfelter K. T., S.M. Boker, J.R. Waddell, N. Vitanov
Journal of Experimental Psychology: Human Perception and Performance
{\bf 35}, 2009, 1072-1091.\\
$[^9]$ Kantz H., D. Holstein, M. Ragwitz, N. K. Vitanov, Physica A 
{\bf 342}, 2004, Nos. 1-2, 315-321.\\
$[^10]$ Puu T., A. Panchuk, Nonlinear economic dynamics. Springer, 
Berlin, 1991.\\
$[^11]$ Vitanov, N. K., E. Yankulova, Chaos, solitons \& Fractals {\bf 28},
2006, No.3, 768-775.\\
$[^{12}]$ Martinov N. K., N. K. Vitanov, J. Phys. A: Math. Gen {\bf 27}, 1994,
No.13, 4611-4618.\\
$[^{13}]$ Thomson J. M. T., H. B. Stewart, Nonlinear dynamics and chaos:
Geometrical methods for scientists. Wiley, New York, 1986.\\
$[^{14}]$ Vitanov N. K. (2000). Eur. Phys. J.  B {\bf 15}, 2000, 
No.2,  349 - 355.\\
$[^{15}]$ Durbin J. S., J. Koopman, Time series analysis by state space methods.
Oxford University Press, Oxford, 2012.\\ 
$[^{16}]$ Panchev S., T. Spassova, N. K. Vitanov N. K. Chaos Solitons 
\& Fractals {\bf 33}, 2007, No.5, 1658-1671.\\
$[^{17}]$ Vitanov N.K., K. Sakai, I. P.  Jordanov, S. Managi, 
K. Demura, Physica A {\bf 382}, 2007, No.1, 330-335 \\
$[^{18}]$ Vespignani A, Science {\bf 325}, 2009, 425-428.\\ 
$[^{19}]$ Vitanov N.K., M. Ausloos, G. Rotundo, Advances in 
Complex Systems {\bf 15}, 2012 Supplement 1, Article Number 1250049.\\
$[^{20}]$ Gai P., S. Kapadia,  Proc. Roy. Soc. London A {\bf 466}, 2010,  
2401 - 2423.\\
$[^{21}]$ Newman M. E. J., Power laws, Contemporary Physics, {\bf 46}, 2005, 
323 - 351.\\
$[^{22}]$ C\' ordoba, J.-C., International  Economic Review, {\bf 49}, 2008,   
1463 - 1468.\\
$[^{23}]$ Zipf, G.K., Human Behavior and the Principle of Least
Effort : An Introduction to Human Ecology. Addison Wesley, Cambridge, MA,
1949.\\
$[^{24}]$ Richardson, H.W., Regional Studies {\bf 7}, 1973, 239-51.\\
$[^{25}]$ Krugman P., Journal of the Japanese and International 
Economies {\bf 10}, 1996, 399-418.\\
$[^{26}]$ Mladenov C., E. Dimitrov,Geography (in Bulgarian) {\bf 1}, 2009, 
13 - 17\\

\begin{flushleft}
$^1$
 G. Nadjakov Institute of Solid State Physics,\\ 
 Bulgarian Academy of Sciences,\\ 
 Blvd. Tzarigradsko Chaussee 72,\\
 BG-1784 Sofia, Bulgaria \\
 e-mail: zdim@issp.bas.bg
\end{flushleft}

\begin{flushleft}
$^2$
GRAPES, Beauvallon Res., rue de la Belle Jardiniere, 483/0021\\
B-4031, Liege Angleur, Euroland  
\end{flushleft}

\end{document}